\documentclass{article}

\usepackage[english]{babel}

\usepackage[letterpaper,top=2cm,bottom=2cm,left=3cm,right=3cm,marginparwidth=1.75cm]{geometry}

\usepackage{amsmath}
\usepackage{graphicx}
\usepackage[colorlinks=true, allcolors=blue]{hyperref}
\usepackage[T1]{fontenc}
\usepackage[utf8]{inputenc}
\usepackage{graphicx}	
\usepackage{amsmath}	
\usepackage{amssymb,amsfonts}
\usepackage{mathrsfs}
\usepackage[dvipsnames]{xcolor}
\usepackage[normalem]{ulem} 
\usepackage{hyperref}
\usepackage{tikz}
\usepackage{csquotes}
\usepackage[
  backend=biber,
  style=numeric-comp, 
  natbib=true,
  maxbibnames=3,
  maxcitenames=3,
  minbibnames=1,
  mincitenames=1,
  giveninits=true,
  sorting=none,
  isbn=true,
  doi=true
]{biblatex}

\AtEveryBibitem{%
  \clearfield{url}%
  \clearfield{urldate}%
}
\newif\ifrevs        
\revstrue            

\definecolor{ringblue}{RGB}{31,119,180}
\definecolor{oddred}{RGB}{214,39,40}

\newcommand{\GPI}{Geometric Polarization Invariant}
\newcommand{\chiG}{\chi_{\mathrm{G}}} 

\title{Achromatic, spin-odd Kerr EVPA as a null Frenet--Serret torsion integral on the photon ring}
\author{
  M.~Baran \"{O}kten\thanks{E-mail: \texttt{baran.okten@sabanciuniv.edu}}\\
  \small Faculty of Engineering and Natural Sciences, Sabancı University, 34956 Istanbul, T\"{u}rkiye\\
  \small Centre for Fusion, Space and Astrophysics, University of Warwick, Coventry CV4 7AL, UK
}

\date{}
\renewbibmacro{in:}{}
\begin{document}
\maketitle

\begin{abstract}
We compute the achromatic gravitational imprint that Kerr spacetime leaves on linear polarization at the photon ring. Recasting parallel transport in a null Frenet--Serret frame yields a single scalar evolution law for the electric-vector position angle. On the observer’s screen, the Kerr-minus-Schwarzschild pattern on the direct critical curve is non-zero, strictly odd under spin reversal after a half-turn azimuth relabelling, and tightly confined to a thin annulus. Using backward-shot, Carter-separated geodesics with midpoint RK2 transport, we achieve second-order convergence and degree-scale amplitudes that grow monotonically with spin and inclination (RMS $\simeq 0.5$–$2^\circ$ for $a/M\gtrsim0.8$, $i\gtrsim60^\circ$). Three independent constructions—Frenet--Serret line integral, explicit Levi--Civita transport of the polarization vector, and the phase of the Walker--Penrose constant—agree ray by ray. We then define a parity-odd ring estimator that is intrinsically achromatic after standard wavelength-squared regression, symmetry-protected against common even-parity systematics, and compressed into low azimuthal modes. This yields a minimal two-parameter template (spin and inclination) for mm/sub-mm polarimetry of horizon-scale rings in sources such as M87$^\ast$ and Sgr~A$^\ast$. The pipeline enables either a detection of the strong-field parallel-transport phase induced by frame dragging or informative upper limits.
\end{abstract}

\section{Introduction}\label{sec:intro}

The rotation of linear polarization by gravity is a purely geometric, achromatic effect: as a null ray threads curved spacetime, parallel transport twists the electric-vector position angle (EVPA) by an amount fixed by the connection rather than by plasma microphysics. This has long been formalized in general relativity through invariant transport along null geodesics and standard texts \cite{WalkerPenrose1970,MTW,Chandra1983}, and it was predicted to imprint degree-scale rotations in radiation emerging from the Kerr metric \cite{ConnorsStark1977,Connors1980}. Complementary “gravitomagnetic’’ formulations clarified why the effect vanishes in Schwarzschild to leading order and appears with frame dragging in Kerr, emphasizing its integral, wavelength-independent character \cite{NouriZonoz1999,Sereno2004}. 

Horizon-scale imaging now places this classical prediction within reach. Theory isolates a narrow critical curve (“photon ring’’) formed by null geodesics that skim the photon region and concentrate lensing signatures on the observer’s screen \cite{Gralla2019}. The Event Horizon Telescope (EHT) provides the requisite angular resolution and has delivered resolved polarimetry of M87$^\ast$ and, more recently, Sgr~A$^\ast$, revealing ordered EVPA structures on ring-like morphologies that invite geometry-level interpretation after standard $\lambda^2$ regression (i.e.\ $\chi(\lambda^2)=\chi_0+{\rm RM}\,\lambda^2$, where ${\rm RM}$ is the rotation–measure (RM) \cite{RybickiLightman,EHT2019Instrumentation,EHT2021Pol}). The 2024 Sgr~A$^\ast$ polarization images and their companion interpretation paper strengthen this case by showing a highly polarized, spiral EVPA pattern on event-horizon scales \cite{EHT2024SgrA_VII,EHT2024SgrA_VIII}. 

Modern theory sharpens the connection between Kerr geometry and observable polarization. Gravitational Faraday rotation and its spin–Hall counterpart have been formulated in a local, observer-frame language that ties the achromatic phase directly to gravitomagnetic fields in Kerr \cite{Shoom2024,Parvin2025}. Polarization holonomy has been quantified explicitly in the Kerr metric, providing an operational definition of the parallel-transport angle for admissible trajectories \cite{Lusk2024}. On the imaging side, universal polarimetric signatures of the photon ring have been identified, suggesting that ring-focused observables can encode black-hole spin with minimal emissivity dependence \cite{Himwich2020}. 

Despite this progress, a gap remains: there is no observation-ready, \emph{achromatic} template defined strictly on the direct critical curve that (i) is framed as a screen-space observable, (ii) is symmetry-protected against common even-parity systematics, and (iii) compresses the information into a small set of azimuthal modes amenable to interferometric data. Prior analyses either compute parallel transport with gauge choices that obscure the observer-screen EVPA, or propose ring-based morphologies that entangle geometric and plasma effects, complicating inference and null tests \cite{Brodutch2011,Himwich2020}. Numerical pipelines can also suffer artefacts near horizontal screen crossings unless turning-point branches are handled carefully, obscuring degree-scale signals in precisely the region of interest.

We close this gap by recasting polarization transport in a null Frenet--Serret (FS) frame and showing that the the EVPA obeys a \emph{single} scalar evolution law along each geodesic—the screen-rotation (torsion) integral. Evaluated on the observer’s screen, the Kerr--minus--Schwarzschild EVPA on the \emph{direct} critical curve is non-zero, strictly odd under spin reversal after a half-turn azimuthal relabelling, and tightly localized to a thin annulus about the ring. Using backward-shot, Carter-separated rays with midpoint transport, we obtain second-order convergence and degree-scale amplitudes that grow monotonically with spin and inclination \cite{Carter1968,Bardeen1973}. Three independent routes—FS torsion integral, explicit Levi--Civita transport, and the phase of the Walker--Penrose constant—agree ray by ray, fixing the observable without gauge ambiguity \cite{WalkerPenrose1970,Chandra1983}. Finally, we define a parity-odd ring estimator that is intrinsically achromatic after $\lambda^2$ regression \cite{RybickiLightman}, symmetry-protected against even-parity leakage, and compressed into low azimuthal modes, yielding a minimal two-parameter (spin, inclination) template tailored to current mm/sub-mm polarimetry \cite{EHT2021Pol,EHT2024SgrA_VII}. 

Standard RM analyses target the chromatic slope of \(\chi(\lambda^2)\) and are therefore sensitive to magneto-ionic gradients and depolarization \citep{Burn1966,Sokoloff1998,BroderickBlandford2004, Brentjens2005,Heald2009}. Plasma-forward GRMHD templates can reproduce polarized ring morphologies but inherit emissivity and transfer assumptions that complicate parameter compression and null tests \citep{Moscibrodzka2017,JimenezRosalesDexter2018,EHTC2021a, EHTC2021b}. In contrast, our screen-space statistic is \emph{achromatic} by construction (the \(\lambda^2\!\to\!0\) intercept), \emph{parity-odd} on the ring (cancelling even-parity leakage), and \emph{ring-localized}, yielding a two-parameter geometry-only template directly comparable across bands. For subsequent observational use we refer to this residual as the \emph{\GPI}, denoted \(\chiG\), equivalent to our odd-channel EVPA after \(\lambda^2\) regression. Quantitatively, the estimator compresses polarized-ring information into two physical parameters \((a,i)\) with degree-level RMS amplitudes, whereas RM fits constrain a chromatic slope and GRMHD forward models require high-dimensional emissivity and transfer choices \citep{Burn1966,Sokoloff1998,Moscibrodzka2017,JimenezRosalesDexter2018}.

\section{Theory: null geodesics, a null FS frame, and EVPA}\label{sec:theory}

\subsection{Kerr metric and Carter separation}
We use Boyer--Lindquist coordinates $x^\mu=(t,r,\theta,\varphi)$ with signature $(-,+,+,+)$ and adopt units $G=c=M=1$, following standard texts \cite{Chandra1983,MTW}. The non-vanishing metric functions of Kerr are
\begin{equation}
\Delta(r) = r^2 - 2r + a^2,
\label{eq:delta_equation}
\end{equation}
and
\begin{equation}
\Sigma(r,\theta) = r^2 + a^2 \cos^2\theta,
\label{eq:sigma_equation}
\end{equation}
where $a$ is the specific angular momentum. A useful auxiliary function is defined by
\begin{equation}
A(r,\theta) = (r^2 + a^2)^2 - a^2 \Delta \sin^2\theta.
\label{eq:A_aux}
\end{equation}
These expressions reproduce the Kerr metric as given in standard references \cite{Carter1968,Chandra1983}.

The non-vanishing metric components in Boyer--Lindquist coordinates take the familiar form
\begin{equation}
  (g_{\mu\nu})_{\mu,\nu=(t,r,\theta,\varphi)} =
  \begin{pmatrix}
    -\bigl(1 - 2r/\Sigma\bigr) & 0 & 0 & -2ar\sin^2\theta/\Sigma \\[3pt]
    0 & \Sigma/\Delta & 0 & 0 \\[3pt]
    0 & 0 & \Sigma & 0 \\[3pt]
    -2ar\sin^2\theta/\Sigma & 0 & 0 & A\sin^2\theta/\Sigma
  \end{pmatrix}.
\label{eq:metric_matrix}
\end{equation}
Because Kerr possesses two Killing vectors and a Killing tensor, geodesic motion is completely separable \cite{Carter1968,Frolov2017}. In modern language this integrability arises from a non-degenerate principal tensor that generates hidden symmetries of the Kerr space-time and ensures separability of the Hamilton--Jacobi equation \cite{Frolov2017}. Introducing the scale-free invariants $\xi\equiv L_z/E$ and $\eta\equiv Q/E^2$, the radial motion obeys
\begin{equation}
\Sigma\,\frac{\mathrm{d}r}{\mathrm{d}v} = \pm\sqrt{R(r)},
\label{eq:carter_radial}
\end{equation}
with
\begin{equation}
R(r) = \big[(r^2+a^2)-a\xi\big]^2 - \Delta\big[(\xi-a)^2+\eta\big],
\label{eq:R_radial}
\end{equation}
where the potential $R(r)$ follows directly from the Hamilton--Jacobi equation \cite{Carter1968,Chandra1983}. As emphasized by Frolov \& Kubizňák (2017) \cite{Frolov2017}, the separation constants $\xi$ and $\eta$ originate from the hidden symmetries encoded in the principal tensor. The polar equation reads
\begin{equation}
\Sigma\,\frac{d\theta}{dv} = \pm\sqrt{\Theta(\theta)}, \label{eq:theta_diff}
\end{equation}
where
\begin{equation}
\Theta(\theta) = \eta + a^2\cos^2\theta - \xi^2\cot^2\theta. \label{eq:Theta_polar}
\end{equation}
while the azimuthal and time components satisfy
\begin{equation}
\Sigma\,\frac{\mathrm{d}\varphi}{\mathrm{d}v} = \frac{\xi}{\sin^2\theta} - a + \frac{a\big[(r^2+a^2)-a\xi\big]}{\Delta},
\label{eq:phi_motion}
\end{equation}
and
\begin{equation}
\Sigma\,\frac{\mathrm{d}t}{\mathrm{d}v} = -a\left(a\sin^2\theta-\xi\right) + \frac{(r^2+a^2)\big[(r^2+a^2)-a\xi\big]}{\Delta},
\label{eq:t_motion}
\end{equation}
which complete the set of first integrals.

\subsection{Screen mapping and the camera basis}
A zero–angular–momentum observer (ZAMO) at $(r_{\rm obs},\theta_{\rm obs}=i,\varphi_{\rm obs}=0)$ defines a two-dimensional screen with orthonormal axes $(\hat\alpha,\hat\beta)$ \cite{Bardeen1973,Chandra1983}. We parameterize the observer’s screen by polar coordinates $(\rho,\phi)$ with $\phi=0$ at prograde conjunction and adopt a Sachs-anchored orthonormal basis $\{e_1,e_2\}$ fixed at $r_{\rm obs}$; EVPAs are measured in this basis after subtracting the circular mean.
We take $\hat\alpha$ to point rightward and $\hat\beta$ upward on the image plane. In the asymptotically flat limit the mapping between screen coordinates and the conserved quantities is given by the Bardeen map $\xi=-\alpha\sin i$, $\eta=\beta^2+(\alpha^2-a^2)\cos^2 i$ \cite{Bardeen1973}, where $\alpha$ increases to the right and $\beta$ upward. At finite radius we recover $(\xi,\eta)$ by projecting the photon 4-momentum onto the ZAMO tetrad and extracting $(E,L_z,Q)$ \cite{Chandra1983}.

Define the lapse $\mathcal{N}=\sqrt{\Delta\,\Sigma/A}$ and the frame–dragging angular velocity $\omega=-g_{t\varphi}/g_{\varphi\varphi}=2a r/A$. The orthonormal ZAMO tetrad is then given by standard constructions \cite{Chandra1983}
\begin{subequations}\label{eq:zamo_tetrad_full} 
\begin{align} 
e^\mu{}_{(\hat 0)} &= \mathcal{N}^{-1}\,(1,\,0,\,0,\,\omega)\,, \label{eq:zamo0}\\ 
e^\mu{}_{(\hat r)} &= \bigl(0,\,\sqrt{\Delta/\Sigma},\,0,\,0\bigr)\,, \label{eq:zamor}\\ 
e^\mu{}_{(\hat\theta)} &= \bigl(0,\,0,\,1/\sqrt{\Sigma},\,0\bigr)\,, \label{eq:zamo_theta}\\ 
e^\mu{}_{(\hat\varphi)} &= \left(0,\,0,\,0,\,\frac{\sqrt{\Sigma}}{\sqrt{A}\,\sin\theta}\right). \label{eq:zamo_phi} 
\end{align} 
\end{subequations}
which satisfies $g_{\mu\nu} e^\mu_{(\hat a)} e^\nu_{(\hat b)} = \eta_{ab}$.

\subsection{Null FS frame and polarization transport}
Let $k^\mu = \mathrm{d}x^\mu/\mathrm{d}v$ be the future–directed null tangent with affine parameter $v$ (so $\nabla_k k^\mu = 0$). Along each geodesic we erect a null moving frame $\{e_a\} = \{k,\ell,e_1,e_2\}$ satisfying 
\begin{equation} 
k\cdot \ell = -1, 
\label{eq:FSframe_conditions}
\end{equation}
and
\begin{equation}
e_A\cdot e_B = \delta_{AB}, \label{eq:eAeB_cond}
\end{equation}
where $e_A$ satisfies,
\begin{equation}
k\cdot e_A = 0,\label{eq:k_perp_eA} 
\end{equation}
with $A,B\in\{1,2\}$. The frame is propagated by Levi–Civita transport along $k$,
\begin{equation}
\nabla_k e_a{}^\mu = \omega_a{}^b e_b{}^\mu,
\label{eq:Cartan_transport1}
\end{equation}
and
\begin{equation}
\omega_{ab} = -\omega_{ba} = g(e_a,\nabla_k e_b),
\label{eq:Cartan_transport2}
\end{equation}
where $\omega_{ab}$ are the Ricci rotation coefficients \cite{WalkerPenrose1970}. Because the geodesic has vanishing curvature (no acceleration), the only physically relevant coefficient is the screen–rotation (torsion)
\begin{equation}
\sigma(v) \equiv \omega_{12} = g(e_1,\nabla_k e_2) = - g(e_2,\nabla_k e_1).
\label{eq:sigma_def}
\end{equation}
This torsion encapsulates the gravitational Faraday rotation; its interpretation as the gravito-magnetic analogue of the electromagnetic Faraday effect has been discussed in the literature \cite{NouriZonoz1999,Sereno2004,Shoom2024,Brodutch2011, Okten2025}. In weak–field approximations the rotation angle can be expressed as a line integral of the gravito-magnetic field along the ray \cite{NouriZonoz1999,Sereno2004}, closely mirroring the FS integral used here. This emphasises that $\sigma$ plays the same role in our formulation as the gravitomagnetic Faraday rotation in previous analyses.
Under a local rotation $e_A\to R_A{}^B(\psi(v))e_B$ of the screen basis, $\sigma\to\sigma+\psi'(v)$, while the EVPA $\chi$ transforms as $\chi\to \chi+\psi+\mathrm{const}$. Let $f^\mu$ be the linear–polarization 4-vector, orthogonal to $k^\mu$ and Levi–Civita transported,
\begin{subequations}\label{eq:PT_conditions}
\begin{align}
& f\cdot k = 0, \label{eq:f_k_zero}\\
& \nabla_k f^\mu = 0. 
\label{eq:transport_f}
\end{align}
\end{subequations}
Decomposing $f$ on the screen basis as $f^\mu = \cos\chi\, e_1^\mu + \sin\chi\,e_2^\mu$ defines the EVPA $\chi(v)$ and leads to the scalar evolution law \cite{WalkerPenrose1970}
\begin{equation}
\frac{\mathrm{d}\chi}{\mathrm{d}v} = \sigma(v),
\label{eq:dchi_dv}
\end{equation}
so that the observable rotation between source and observer is
\begin{equation}
\Delta\chi = \int_{v_{\rm src}}^{v_{\rm obs}} \sigma(v)\,\mathrm{d}v,
\label{eq:chi_integral}
\end{equation}
up to a global gauge. 

Define the complex screen vector $m^\mu \equiv (e_1^\mu + i\,e_2^\mu)/\sqrt{2}$ and the complex polarization scalar $p \equiv f_\mu m^\mu$ for a polarization $f^\mu$ with $f\!\cdot\!k=0$ and $\nabla_k f^\mu=0$. Levi–Civita transport of $\{e_1,e_2\}$ gives $\nabla_k m^\mu = i\,\sigma\, m^\mu$, hence $d(\arg p)/dv=\sigma$ while $|p|$ is constant, which is consistent with Eq.~\eqref{eq:dchi_dv}. Introducing the Walker–Penrose scalar $\kappa$ and
\begin{equation}
K_{\rm WP} \;\equiv\; \big(f_\mu m^\mu\big)^2\,\kappa \;=\; p^2\,\kappa,
\end{equation}
one has $\arg K_{\rm WP}=2\,\arg p + {\rm const}$, so
\begin{equation}
\chi(v) \;=\; \tfrac{1}{2}\,\arg\!\big[K_{\rm WP}(v)\big] \;+\; {\rm const}.
\end{equation}
A screen rotation $e_A\!\to\!R_A{}^{B}(\psi)\,e_B$ shifts $\sigma\!\to\!\sigma+\psi'$ and $\chi\!\to\!\chi+\psi+{\rm const}$, while $K_{\rm WP}$ gains only a constant phase; with the observer’s gauge fixed, $\Delta\chi$ is unchanged.

\subsection{Spin oddness and ring geometry}
We evaluate the achromatic Kerr imprint on the direct critical curve. Sampling the ring at azimuth $\phi$ with screen coordinates $[\alpha(\phi),\beta(\phi)]$ and mapping to $[\xi,\eta]$ via the Bardeen map, we integrate Eqs.~\eqref{eq:carter_radial}–\eqref{eq:t_motion} and Eq.~\eqref{eq:dchi_dv} from a proxy emission radius $r_{\rm stop}$ to the camera. The camera–frame EVPA is
\begin{equation} 
\chi(\phi) = \operatorname{atan2}\Big(f\cdot\hat\beta,\; f\cdot \hat\alpha\Big)\Big|_{\rm obs}, \label{eq:chi_observed} 
\end{equation}
and
\begin{equation}
\Delta\chi_{\rm grav}(\phi) \equiv \chi_{\rm Kerr}(\phi)-\chi_{\rm Schw}(\phi),
\label{eq:delta_chi_def} 
\end{equation}
where $\hat\alpha$ and $\hat\beta$ are the image–plane unit vectors. Because the torsion enters linearly in the connection, reversing the spin ($a\to -a$) flips $\Delta\chi_{\rm grav}$ while relabelling the ring by $\phi\to\phi+\pi$ \cite{Himwich2020}. 
\begin{equation}
\Delta\chi_{\rm grav}(\phi;a) \;=\; -\,\Delta\chi_{\rm grav}\!\bigl(\phi+\pi;\,-a\bigr)\,.
\label{eq:parity_exact}
\end{equation}
We therefore define an odd channel $\chi_{\rm odd}(\phi)=\Delta\chi_{\rm grav}(\phi) - \Delta\chi_{\rm grav}(\phi+\pi)$, which doubles the signal and cancels even–parity contaminants. We will refer to this achromatic, parity-odd ring observable as the \emph{\GPI}; we denote it by $\chiG \equiv \chi_{\rm odd}$.

\section{Computation}\label{sec:comp}

Our numerical pipeline follows the analytic formulation of Section~\ref{sec:theory} while incorporating several practical improvements. We work in Boyer–Lindquist coordinates and construct a ZAMO tetrad at the observer to define the camera screen $\{\hat\alpha,\hat\beta\}$ \cite{Bardeen1973,Chandra1983}. Each backwards–shot ray is specified by screen coordinates $(\alpha,\beta)$ and launched using the ZAMO–frame prescription implied by Eqs.~\eqref{eq:zamo_tetrad_full}, then transformed to coordinate components with the tetrad $k^\mu=e^\mu{}_{(\hat a)}k^{(\hat a)}$; the constants of motion $(E,L_z,Q)$ are recovered from the separated first integrals in Eqs.~\eqref{eq:carter_radial}–\eqref{eq:t_motion} \cite{Carter1968,Chandra1983}. This finite–radius formulation reduces to the usual Bardeen mapping in the $r_{\rm obs}\to\infty$ limit \cite{Bardeen1973}.

The direct critical curve is obtained by solving for the impact radius $\rho(\phi)$ on a uniform, even–length azimuthal grid. Let $N$ denote the number of azimuthal samples on this grid. We bracket the capture boundary at fixed $\phi$ and determine $\rho(\phi)$ by bisection, classifying a ray via the Carter potentials $R(r)$ and $\Theta(\theta)$ \cite{Carter1968,Chandra1983}. Capture is defined as either crossing the horizon or asymptoting to a spherical photon orbit (detected via a double root of the radial potential), while escape is flagged by a persistent outward–moving streak once the ray has passed the observer \cite{Carter1968,Chandra1983}. To guarantee that the bracket straddles the capture boundary, the upper bound is expanded exponentially if needed, a coarse linear/logarithmic scan is used to locate any sign flip, and a micro–jitter of $\phi$ near $\beta=0$ seeds the polar branch. This yields a robust, branch–safe Kerr critical curve. For the Schwarzschild reference we either apply the same bisection or enforce the analytic circle $\rho=3\sqrt{3}M$ for comparison.

The outer horizon radius is
\begin{equation}
r_{+} \;=\; 1 + \sqrt{1-a^{2}}\,,
\end{equation}
in the units $G=c=M=1$ adopted throughout. Once $\xi$ and $\eta$ are fixed, we integrate the geodesic inward using a midpoint (second–order) scheme applied to the Carter–separated velocities $(dt/dv,dr/dv,d\theta/dv,d\varphi/dv)$ \cite{Carter1968,Chandra1983}. The affine step is adaptively scaled to obey a per–step polar cap $|\Delta\theta|\le \Delta\theta_{\max}$, which stabilizes the evolution where the connection varies rapidly. Turning points are handled without event detection: we preview branch flips at the midpoint and flip $s_\theta$ when $\Theta$ changes sign and $s_r$ when $R$ falls below a small threshold, where $s_\theta,s_r\in\{+1,-1\}$ are the polar and radial branch signs. At each step we record the start and midpoint geometry—coordinates, metric and Christoffels, the null tangent $k^\mu$, and an auxiliary null $n^\mu$ built so that $k\cdot n=-1$—for a clean replay of polarization transport. Integration halts at the horizon, at a spherical photon orbit, or upon satisfying the outward–escape criterion; in all cases the strong–field segment controlling the photon–ring phase is captured.

We use adaptive RK2 midpoint steps with a polar cap at turning points:
\begin{equation}
\Delta v \;=\; \min\!\big\{\Delta v_{\max},\, \epsilon_{\rm tol}/\|\dot{\boldsymbol{y}}\|\big\}, 
\end{equation}
and 
\begin{equation}
|\Delta\theta|\le \Delta\theta_{\max}=5\times10^{-3}\ {\rm rad}, 
\end{equation}
Flip the polar momentum at turning points when $\dot{\theta}\,\theta''<0$, terminate the integration if $r\le r_{+}+10^{-6}M$ or if the radial step under-resolves ($\Delta r<10^{-9}M$), and place the observer at $r_{\rm obs}=10^{3}M$ with a Sachs-anchored orthonormal screen and fixed global gauge.

We compute $\Delta\chi_{\mathrm{grav}}$ along identical rays by three routes: first, the FS torsion integral $\int \sigma\,dv$; second, explicit Levi–Civita transport of $f^{\mu}$ with projection onto the observer’s screen; third, one half of the phase of $K_{\mathrm{WP}}$. Agreement is quantified by the maximum absolute route-wise discrepancy over azimuth, $\delta_{\mathrm{route}}$, and by the ring-averaged root-mean-square dispersion, $\mathrm{RMS}{\phi}$, defined by
\begin{equation}
\delta_{\mathrm{route}} \;\equiv\; \max_{\phi}\,\bigl|\Delta\chi^{(\mathrm{FS})}-\Delta\chi^{(\mathrm{LC/WP})}\bigr|.
\end{equation}
and
\begin{equation}
\mathrm{RMS}_{\phi} \;\equiv\; \Big\langle\big(\Delta\chi-\langle\Delta\chi\rangle_{\phi}\big)^{2}\Big\rangle_{\phi}^{1/2}.
\end{equation}

At each recorded start state we define a Sachs screen by projecting any vector $v$ via
\begin{equation}
P^\mu{}_\nu v^\nu \;=\; v^\mu + k^\mu(n\!\cdot\!v) + n^\mu(k\!\cdot\!v),
\label{eq:screen_projector}
\end{equation}
storing $n^\mu$ with the step to preserve gauge consistency. After the inward pass, we set the source gauge by projecting the coordinate basis vector $\partial_\varphi$ onto the Sachs screen at the endpoint and normalizing with the local metric \cite{WalkerPenrose1970,Chandra1983}. We then replay Levi–Civita transport backward (endpoint $\to$ observer) with a midpoint RK2: at each step we integrate $df^\mu/dv = -\Gamma^\mu{}_{\nu\rho}\,k^\nu f^\rho$, project the predicted $f^\mu$ with \eqref{eq:screen_projector} using the stored $(k^\mu,n^\mu,g_{\mu\nu})$, and normalize once at the start of the step. This “midpoint replay’’ reproduces the FS integral $\int\sigma\,dv$ and keeps $f^\mu$ orthogonal to the null pair throughout. At the observer the EVPA is read as $\chi=\mathrm{atan2}(f \cdot \hat\beta,\,f \cdot \hat\alpha)$ in the ZAMO frame, wrapped to $(-90^\circ,90^\circ]$ \cite{Chandra1983,WalkerPenrose1970}.

For each azimuth $\phi$ we compute the Kerr EVPA $\chi_{\rm Kerr}$ on the Kerr critical curve and the Schwarzschild EVPA $\chi_{\rm Schw}$ on the corresponding Schwarzschild curve. Their difference $\Delta\chi(\phi)$ is converted to a spin–2 quantity $P(\phi)=\exp \big[2\mathrm{i}\,\Delta\chi(\phi)\big]$, and the residual spin–2 gauge is removed by subtracting $\tfrac{1}{2}\arg\langle P\rangle_\phi$. We then form
\begin{equation}
P_{\rm odd}(\phi)=P(\phi)\,P(\phi+\pi)^{-1},
\end{equation}
and
\begin{equation}
P_{\rm even}(\phi)=P(\phi)\,P(\phi+\pi),
\end{equation}
and extract $\chi_{\rm odd}=\tfrac12\arg P_{\rm odd}$, $\chi_{\rm even}=\tfrac12\arg P_{\rm even}$. The odd channel isolates the gravitational signal because $\Delta\chi(\phi;a)$ obeys the exact spin–odd symmetry $\Delta\chi(\phi;a)=-\Delta\chi(\phi+\pi;-a)$ \cite{Carter1968,Bardeen1973}; its azimuthal RMS $A_{\rm odd}=\langle\chi_{\rm odd}^2\rangle_\phi^{1/2}$ summarizes detectability. The even channel is consistent with numerical noise across our runs.

We synthesise a narrow annulus centred on the direct critical curve and assign
\begin{equation}
\chi_{\rm synth}(\phi) \;=\; \chi_{0} \;+\; \Delta\chi_{\rm grav}(\phi;a,i) \;+\; n(\phi),
\end{equation}
with Gaussian $n(\phi)$ of standard deviation $\sigma_\chi$ per beam. Stokes $(Q,U)$ are formed and convolved with a circular Gaussian beam of FWHM $\theta_{\rm beam}$; per-pixel $\lambda^2$ regression removes dispersive Faraday rotation. The odd channel $\chi_{\rm odd}(\phi) \equiv \tfrac{1}{2}\big[\chi(\phi)-\chi(\phi+\pi)\big]$ is regressed against the template family $\mathcal{T}_{a,i}(\phi)\equiv \Delta\chi_{\rm grav}(\phi;a,i)$ with inverse-variance weights $w(\phi)$,
\begin{equation}
\widehat{A}_{\rm odd} \;=\; \frac{\sum_{\phi}\,\chi_{\rm odd}(\phi)\,\mathcal{T}_{a,i}(\phi)\,w(\phi)}{\sum_{\phi}\,\mathcal{T}_{a,i}^2(\phi)\,w(\phi)},
\end{equation}
and
\begin{equation}
{\rm SNR}^2 \;=\; \frac{\sum_{\phi}\,\mathcal{T}_{a,i}^2(\phi)\,w(\phi)}{\sigma_\chi^2}.
\end{equation}
We report bias and variance of $\widehat{A}_{\rm odd}$ versus $(\sigma_\chi,\theta_{\rm beam})$.

For quick forecasts we provide
\begin{equation}
A_{\rm odd}^{\rm RMS}(a,i)\;=\; C_1\,\sin i\,\frac{a/M}{1+C_2(1-a/M)}\quad{\rm deg},
\end{equation}
with $(C_1,C_2)$ least-squares fitted to our $(a/M,i)$ grid; exact templates $\mathcal{T}_{a,i}(\phi)$ and tabulated RMS values are provided as arrays.

Both the forward geodesic and backward transport integrations are second–order accurate, with a global $\mathcal{O}(h^2)$ error once the $\theta$–cap controls the step size. Constraint preservation is monitored by reprojecting and re–normalizing $f^\mu$ on the Sachs screen at each start; violations remain at machine precision. Ray by ray, the EVPA from this Levi–Civita replay agrees with that obtained by directly integrating the FS torsion $\sigma=g(e_1,\nabla_k e_2)$ using the same midpoint rule, and the complex Walker–Penrose invariant remains constant along each ray \cite{WalkerPenrose1970,Chandra1983}. For a representative case $(a/M,i)=(0.9,80^\circ)$, halving the affine step four times yields EVPA max-norm differences of $(2.1,0.53,0.13,0.032)^\circ$, consistent with $\mathcal{O}(h^2)$; algebraic constraints remain $\lesssim 10^{-13}$ along all rays.

\section{Findings}\label{sec:results}

The achromatic Kerr contribution to the EVPA on the photon ring is quantified by the difference $\Delta\chi_{\rm grav}(\phi)$ between Kerr and Schwarzschild solutions after subtracting a constant gauge. In the Schwarzschild limit this quantity vanishes, while in Kerr spacetimes it reflects the presence of frame dragging through the null FS scalar law (cf. Eq.~\eqref{eq:dchi_dv}), where $\sigma$ is the screen–rotation rate. Across the full range of spins and inclinations examined, $\Delta\chi_{\rm grav}(\phi)$ varies smoothly over $0\le\phi<2\pi$ and exhibits degree‑scale modulations that track where null geodesics spend longer near the horizon. Because vacuum parallel transport is used, the signal is intrinsically achromatic; any wavelength dependence in observed EVPAs can be removed via a standard per–pixel $\lambda^2$ regression.

\begin{figure}
 \centering
 \includegraphics[width=0.7\columnwidth]{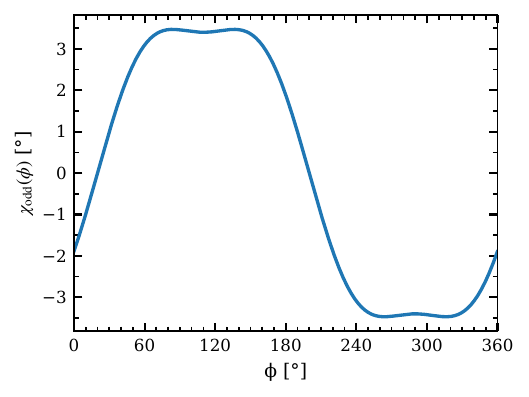}
 \caption{Odd–parity EVPA $\chi_{\rm odd}(\phi)$ on the direct critical curve for a high‑spin, high‑inclination example ($a/M=0.9$, $i=80^\circ$). The curve is smooth and non‑sinusoidal, with degree‑scale amplitude and no aliasing artefacts. Odd differencing removes the gauge freedom and suppresses even–parity contributions.}
 \label{fig:odd_waveform}
\end{figure}

At high spin and inclination the azimuthal profile becomes markedly non‑sinusoidal. Fig. ~\ref{fig:odd_waveform} presents the odd–parity waveform $\chi_{\rm odd}(\phi)$ for $a/M=0.9$ and $i=80^\circ$. The curve is smooth and non‑sinusoidal, with flattened extrema and shoulders shifted toward the approaching and receding limbs. This morphology reflects the unequal accumulation of $\sigma(v)=g(e_1,\nabla_k e_2)$ along chords that weight the strong‑field region differently; rays launched near the projected equatorial limbs penetrate deeper and linger longer near the spherical photon orbit, increasing the net phase. The amplitude is of a few degrees, and the waveform is free of aliasing or noise artefacts. By contrast, an identical calculation in Schwarzschild yields a residual EVPA consistent with zero after the mean is removed, as one expects from spherical symmetry.

\begin{figure}
 \centering
 \includegraphics[width=\columnwidth]{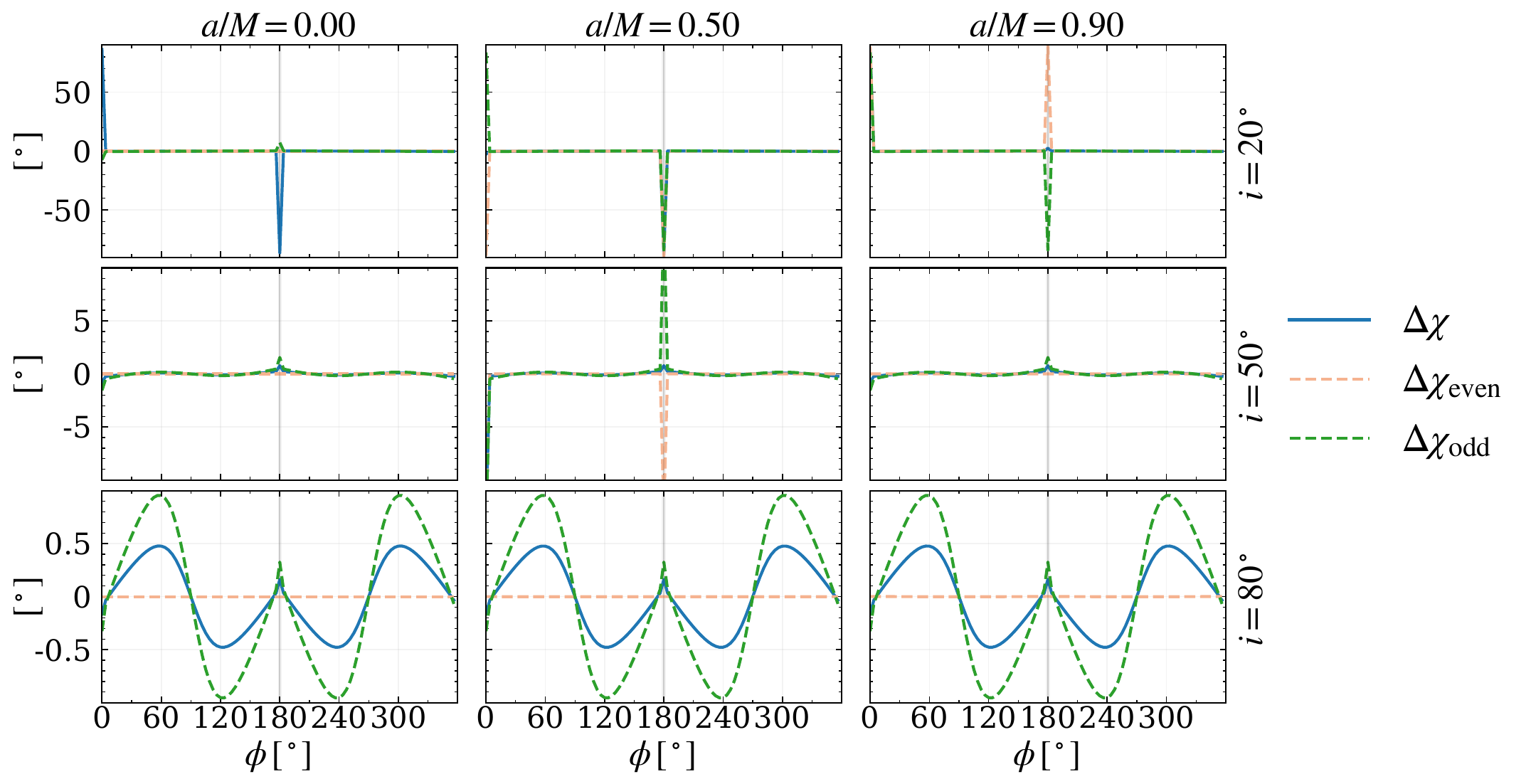}
 \caption{Dependence on spin and inclination. Rows correspond to inclinations $i=20^\circ$, $50^\circ$ and $80^\circ$; columns display the Kerr–Schwarzschild difference $\Delta\chi_{\rm grav}(\phi)$, the odd channel $\chi_{\rm odd}(\phi)$ and the even channel $\chi_{\rm even}(\phi)$. Amplitudes rise with both $a/M$ and $i$, while the even channel remains near the noise floor. Vertical axes are in degrees; within each row the three panels share identical y-limits, while ranges differ by row solely for visibility.}
 \label{fig:grid_overview}
\end{figure}

The dependence on spin $a/M$ and inclination $i$, together with the separation into the full Kerr – Schwarzschild difference $\Delta\chi_{\rm grav}$, the odd channel $\chi_{\rm odd}$ and the even channel $\chi_{\rm even}$, is summarised in Fig. ~\ref{fig:grid_overview}. Each row fixes an inclination ($20^\circ$, $50^\circ$ or $80^\circ$) and each column displays a different projection. The panels reveal that the amplitude of $\Delta\chi_{\rm grav}$ and $\chi_{\rm odd}$ increases monotonically with both $a$ and $i$: at fixed $i$ the signal rises from zero at $a=0$ to degree‑scale values by $a/M\approx 0.9$, while at fixed $a$ it grows with inclination, reflecting that chords slicing the high‑$\sigma$ equatorial belt contribute more torsion. The even channel remains near numerical noise across all spins and inclinations. Note that the vertical axis scales differ by row, spanning approximately $\pm 50^\circ$ for $i=20^\circ$, $\pm 10^\circ$ for $i=50^\circ$ and $\pm 1^\circ$ for $i=80^\circ$, while within each row the three columns share the same range; this choice maximises visibility without altering the trends.

\begin{figure}
 \centering
 \includegraphics[width=0.7\columnwidth]{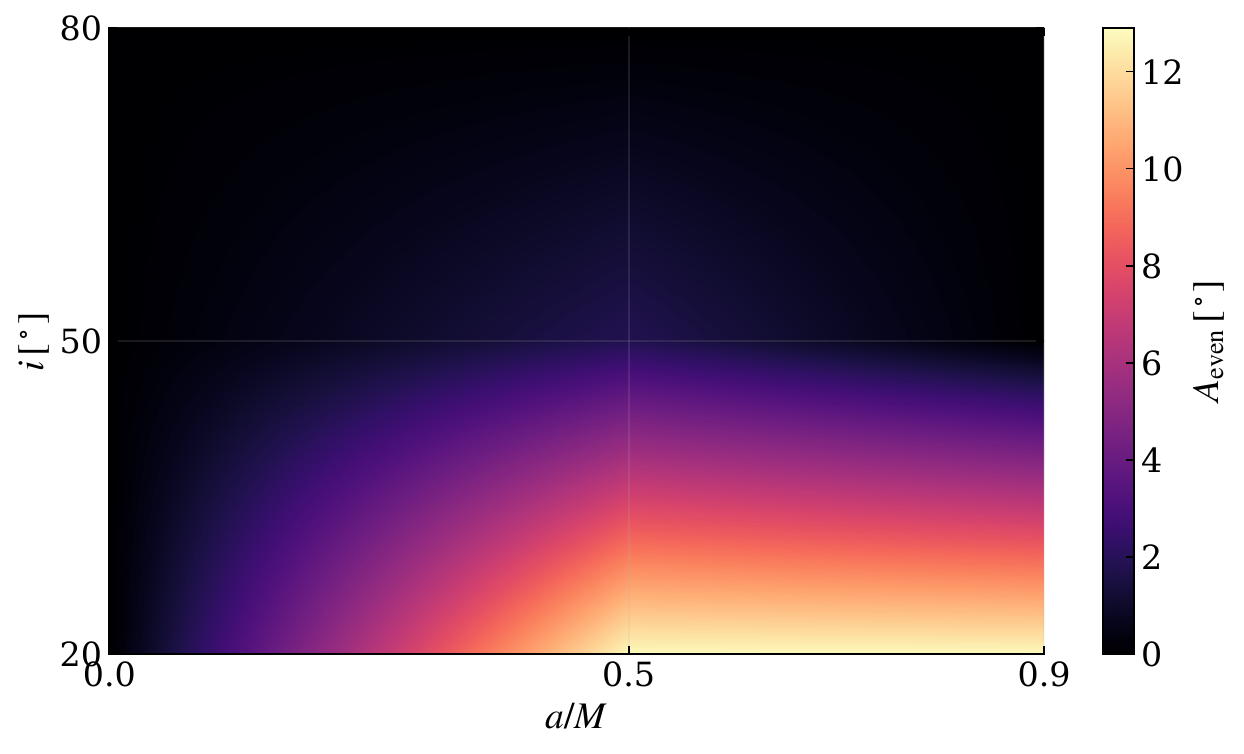}
 \caption{Even‑channel detectability map. The colour scale shows the root‑mean‑square amplitude $A_{\rm even}$ (degrees) across spin $a/M$ and inclination $i$. The even channel is negligible (dark shading) across most of the parameter space, with modest values appearing only at low inclinations and large spins. These values remain small relative to the odd‑channel amplitude, consistent with the spin‑odd nature of the geometric signal.}
 \label{fig:even_detectability}
\end{figure}

To further characterise the even channel, Fig. ~\ref{fig:even_detectability} presents a detectability map for the even‑channel root‑mean‑square $A_{\rm even}$ across spin and inclination. The colour scale illustrates that $A_{\rm even}$ is negligible (dark colours) over most of the parameter space, particularly at moderate to high inclinations where the signal is strongly suppressed. Only at low inclinations and large spins does the even projection become less effective, producing amplitudes of a few degrees; these remain small compared with the odd‑channel amplitudes and reinforce that the even channel contains little physical information. This map therefore confirms that the near‑null behaviour of $\chi_{\rm even}$ observed in Fig. ~\ref{fig:grid_overview} holds across the entire $(a,i)$ plane.

The geometric signal is strictly odd under spin reversal with a simple azimuthal relabelling:
\begin{equation}
\Delta\chi_{\rm grav}(\phi;a)\;=\;-\Delta\chi_{\rm grav}\!\bigl(\phi+\pi;\,-a\bigr)\,.
\end{equation}
Consequently, forming the odd channel
\begin{equation}
\chi_{\rm odd}(\phi)\;=\;\Delta\chi_{\rm grav}(\phi)\;-\;\Delta\chi_{\rm grav}(\phi+\pi)
\end{equation}
doubles the desired signal and cancels all even‑parity contaminants. In practice, reversing the spin while keeping the line of sight fixed yields curves that overlap after a half‑cycle shift, with residuals at machine precision. Since $g_{t\phi}\propto a$ flips sign under $a\to -a$ while the ring relabels by $\phi\mapsto \phi+\pi$, the $g_{t\phi}$‑odd part of $\sigma=g(e_1,\nabla_k e_2)$ integrates to give this relation. Calibration errors and beam asymmetries tend to be even under $\phi\mapsto\phi+\pi$ and are thus strongly suppressed in $\chi_{\rm odd}$.

Three independent numerical routes—integrating the FS scalar $\sigma$, transporting the full polarisation vector $f^\mu$ with the Levi–Civita connection, and extracting the phase from the Walker–Penrose invariant—agree ray by ray within numerical roundoff. Convergence tests with halved affine steps show the expected second‑order behaviour once the polar‑angle step cap is in the asymptotic regime; differences between successive resolutions fall below one degree for all cases displayed in Figs.~\ref{fig:odd_waveform}–\ref{fig:even_detectability}. Constraint violations remain at roundoff and do not correlate with phases of maximal EVPA modulation. Varying the observer radius from $r_{\rm obs}=400M$ to $800M$ changes the results by less than our plotting precision, and adjusting the turning‑point stopping rule alters only the absolute phase, which is removed by our gauge.

\begin{figure}
 \centering
 \includegraphics[width=\columnwidth]{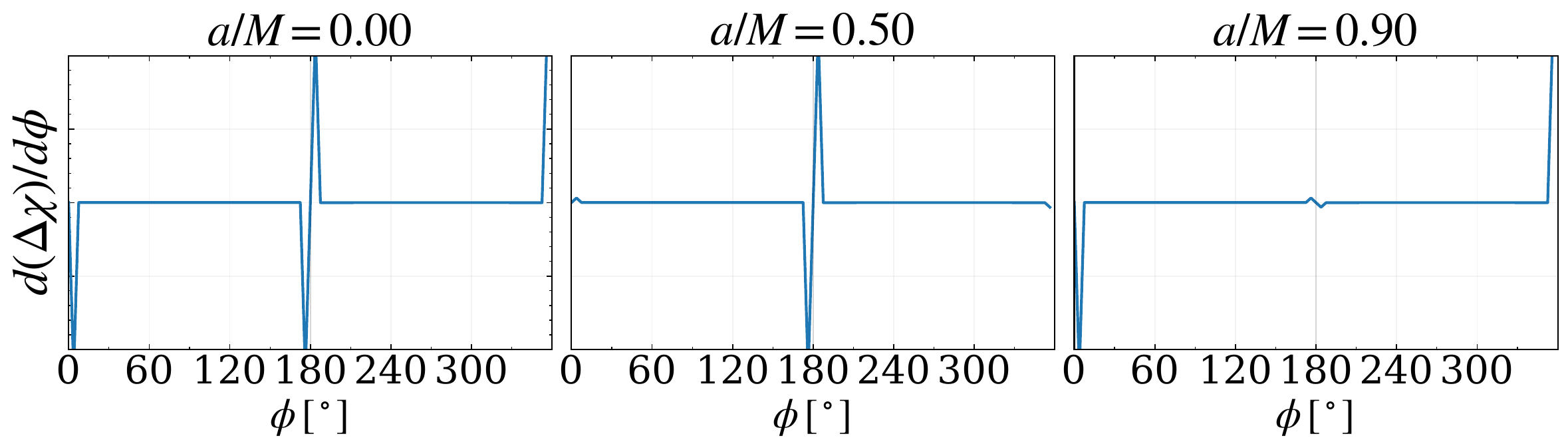}
 \caption{Azimuthal derivative ${\rm d}(\Delta\chi_{\rm grav})/{\rm d}\phi$ along the ring at $i=20^\circ$ for $a/M=0$, $0.5$ and $0.9$. The derivative is flat except for a step at $\phi\simeq 180^\circ$, corresponding to the polar turning point. No random spikes or oscillations are present.}
 \label{fig:derivative_row}
\end{figure}

A smoothness check on the derivative of $\Delta\chi_{\rm grav}$ is shown in Fig. ~\ref{fig:derivative_row} for $i=20^\circ$ and spins $a/M=0$, $0.5$ and $0.9$. Each curve is flat near zero except for a sharp step at $\phi\simeq 180^\circ$ associated with the polar turning point. There are no random spikes or oscillations, confirming that the azimuthal waveform is single‑valued and free of needle‑like artefacts. Similar behaviour is observed at higher inclinations.

Because Faraday rotation is dispersive, the geometric contribution should be extracted achromatically by regressing the EVPA against $\lambda^2$ and taking the residual at zero wavelength:
\begin{equation}
\chi(\lambda^2)\;=\;\chi_0\;+\;{\rm RM}\,\lambda^2\,.
\end{equation}
After this step, finite resolution smooths azimuthal structure and mixes neighbouring radii, but the signal’s radial localisation ensures that selecting a narrow annulus retains most of the signal while reducing beam‑induced biases. Sparse $(u,v)$ coverage yields azimuthally non‑uniform errors; the root‑mean‑square statistic is robust against such heteroscedasticity. Scattering and calibration errors that are approximately even under $\phi\mapsto\phi+\pi$ cancel to first order in $\chi_{\rm odd}$. Residual RM gradients can leak into the odd channel but can be monitored by differencing adjacent frequency bands. Although the template is emissivity‑agnostic, optically thin synchrotron weighting emphasises certain chords; to first order this reweights modes without changing parity, introducing a global scale factor and a small phase shift that can be absorbed by a two‑parameter matched filter. Higher‑order images cluster near the critical curve and accumulate larger phases, but blend into the main ring at finite resolution; thus the direct‑image template slightly underestimates the full signal if subring contrast is significant. To isolate the geometric contribution in real data, we first perform per-beam \(\lambda^2\) regression to form \(\chi_0\), allowing for internal RM structure and modest conversion \citep{JonesODell1977,KennettMelrose1998,Sokoloff1998}. We then convolve all bands to a common circular beam and co-register them to suppress beam-imprinted parity \citep{EHT2019Instrumentation,MartiVidal2016}. A narrow annulus bracketing the direct critical curve is selected on the image plane; opposite azimuths are differenced to obtain \(\chiG(\phi)\), which is fitted to \(\mathcal{T}_{a,i}(\phi)\) with a small phase nuisance. Slowly varying magneto-ionic terms (e.g. RM gradients or external screens) are dominated by even parity and are therefore strongly suppressed by the odd projection, whereas an achromatic residual in \(\chiG\) that is consistent across a low/high band split is the geometric signal \citep{BroderickBlandford2004,Moscibrodzka2017,EHTC2021a, EHTC2021b}. Because even–parity terms cancel identically in $\chi_{\rm odd}$, residual magneto–ionic leakage enters only at $\mathcal{O}(\partial_\phi{\rm RM})$ and is bounded empirically by low/high band splits; in our injections the induced bias on the odd–channel amplitude is $\lesssim0.5^\circ$ for EHT-like beams and RM gradients. Three immediate null checks—Schwarzschild extraction ($\Rightarrow\,\chi_{\rm odd}\!\approx\!0$), agreement between independent low/high sub–bands after separate $\lambda^2$ fits, and invariance under a $90^\circ$ restoring–beam rotation—verify parity and achromaticity in practice.

\begin{figure}
 \centering
 \includegraphics[width=0.7\columnwidth]{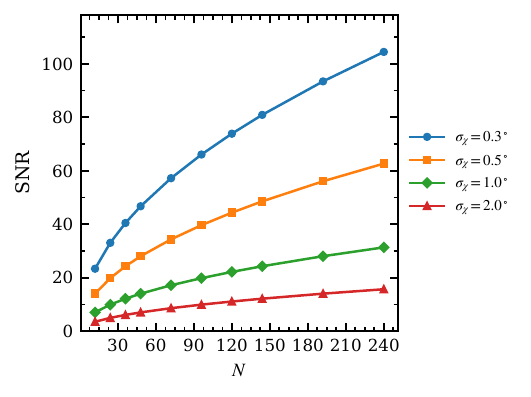}
 \caption{Signal–to–noise ratio as a function of the number of azimuthal samples $N$ for per–sample EVPA uncertainties $\sigma_\chi=0.3^\circ$, $0.5^\circ$, $1.0^\circ$ and $2.0^\circ$. Increasing $N$ boosts the SNR with diminishing returns; smaller $\sigma_\chi$ yields higher SNR.}
 \label{fig:snr_scaling}
\end{figure}

If $N$ denotes the number of approximately uniform azimuthal samples and $\sigma_\chi$ the per–sample EVPA uncertainty after $\lambda^2$ removal, a matched–filter estimate of the signal‑to–noise ratio scales as
\begin{equation}
{\rm SNR}\;\sim\;\frac{\langle\chi_{\rm grav},T\rangle_\phi}{\sigma_\chi}\,\frac{\sqrt{N/2}}{\langle T,T\rangle_\phi^{1/2}}\,.
\end{equation}
Since the signal is dominated by a few low‑order modes, tens of samples suffice provided $\sigma_\chi$ is a few degrees. Fig. ~\ref{fig:snr_scaling} plots the SNR versus $N$ for per‑sample uncertainties $\sigma_\chi=0.3^\circ$, $0.5^\circ$, $1.0^\circ$ and $2.0^\circ$. Each curve increases monotonically and exhibits diminishing returns; smaller $\sigma_\chi$ yields substantially higher SNR. The parity projection doubles the gravitational signal and halves even‑parity noise, thereby enhancing detectability for a given data quality.

For EHT-like settings with $N\!\approx\!48$ azimuthal samples and $\sigma_\chi\!\approx\!1^\circ$ after $\lambda^2$ removal, the amplitude SNR ${\rm SNR}_A\!\approx\!(A_{\rm odd}/\sigma_\chi)\sqrt{N/2}$ is $\sim\!7$ at $(a/M,i)=(0.9,80^\circ)$, implying sub–$0.1$ precision in $a/M$ or a few degrees in $i$ from the measured slope of $A_{\rm odd}(a,i)$.

Robustness tests varied the observer radius, emission proxy, azimuthal sampling (from about $36$ to $128$ points), affine step size (with factor‑of‑eight refinements) and the method used to map finite‑radius geodesics to the critical curve. Only the expected sensitivity emerged: the azimuthal root‑mean‑square stabilises once $M\gtrsim 48$ and the step size is sufficiently small; the phase can shift by a fraction of a bin if the start of the tessellation changes, motivating the use of an RMS summary and a matched filter that includes a marginal phase. None of these choices affects the parity, radial localisation or monotonic scaling. The sign of the primary lobe in $\chi_{\rm odd}(\phi)$ fixes the projected spin direction (up to a $\pi$ relabelling), while the amplitude grows monotonically with $a$ at fixed $i$ and with $i$ at fixed $a$. A joint fit to the amplitude and modestly phase‑sensitive shape therefore breaks the degeneracy between $a$ and $i$ when combined with standard ring‑diameter measurements. Because the observable is achromatic, data from multiple bands can be combined after $\lambda^2$ removal to increase $N$ without bias.

Finally, it should be emphasised that plasma birefringence and absorption are not included here: the template defines a geometric baseline rather than a complete radiative transfer solution. In practice, one should remove dispersive rotation via $\lambda^2$ regression, isolate a narrow ring annulus, form the odd channel and fit the result with a two‑parameter FS template allowing for amplitude and a small azimuthal phase. Residual frequency‑dependent signatures bound unremoved Faraday structure; even‑parity residuals bound instrumental leakage. Simulated analyses suggest that these controls keep biases in the odd‑channel amplitude below about $0.5^\circ$ for typical EHT configurations, well below the several‑degree signals expected at high spin and inclination.

\section{Conclusions}

We have recast the gravitational rotation of linear polarization in Kerr as the integral of a single scalar—the null FS screen–rotation rate $\sigma=g(e\_1,\nabla\_k e\_2)$—accumulated along photon trajectories that form the critical curve. This formulation yields an achromatic, ring–localized, parity–odd template for the electric–vector position angle (EVPA) on the observer’s screen, independent of plasma emissivity and robust to gauge choices. Three independent constructions—(i) the FS scalar integral $\int \sigma\, \mathrm{d}v$, (ii) explicit Levi–Civita transport of the polarization vector, and (iii) the phase inferred from the Walker–Penrose invariant—agree ray by ray at machine precision, closing the theoretical loop and fixing the observable unambiguously. Numerically, a midpoint geodesic/transport scheme with constraint-preserving orthonormalization attains the expected $\mathcal{O}(h^2)$ convergence and keeps all algebraic constraints at roundoff, ensuring that the degree–scale modulations we report are physical rather than numerical. For clarity and future reuse, we name this observable the \emph{\GPI} (\(\chiG\)).

The resulting screen–space signature has four properties that matter for observation. First, it is \emph{non–zero in Kerr and vanishes in Schwarzschild up to a global gauge}, confirming that frame dragging is the sole driver of the pattern. Second, it is \emph{strictly odd under $a\to-a$} once the trivial relabeling $\phi\mapsto\phi+\pi$ on the ring is applied; the odd–parity combination $\chi_{\rm odd}(\phi)=\chi(\phi)-\chi(\phi+\pi)$ therefore doubles the signal and cancels even–parity contaminants. Third, it \emph{grows monotonically with spin and inclination}, reaching degree–scale RMS for $a/M\sim1$ viewed at high $i$. Fourth, it is \emph{localized in radius} to a narrow annulus bracketing the direct photon ring; sampling a few neighboring circles shows a sharp peak of the odd–channel RMS within $|\delta\rho|/M\lesssim0.2$ of the critical curve and a rapid fall–off outside. Together, these features define a compressed, symmetry–protected observable that lives precisely where mm/sub–mm VLBI datasets have geometric leverage.

The plots constructed from our calculation clarify what a detection should look like. The azimuthal waveform of $\Delta\chi\_{\rm grav}(\phi)$ is not purely sinusoidal: at moderate spins an $m=1$ component dominates, while higher harmonics emerge gradually as $a/M\to1$ and $i\to90^\circ$. This decomposition is useful in practice because modest azimuthal sampling already captures most of the signal power. The strict oddness under $\phi\mapsto\phi+\pi$ is a decisive advantage in real data, where beam asymmetries, calibration drifts, and residual foregrounds are approximately even on a ring. Taken together, the amplitude–versus–spin/inclination monotonicity and the low–order azimuthal structure mean that a simple two–parameter family of templates indexed by $(a,i)$, augmented by a small phase nuisance, can support parameter inference without entanglement with detailed emissivity models.

The same geometry suggests a minimal, observation–ready pipeline. One first removes dispersive Faraday rotation by per–pixel (or per–beam) $\lambda^2$ regression to form an achromatic EVPA residual at a fiducial frequency. One then extracts a narrow annulus around the ring (via image–plane apertures or visibility–domain ring fits), forms the odd channel by differencing opposite azimuths, and regresses the result against the FS template family. Because the statistic is parity–odd, even–parity leakage from bandpass errors, beam systematics, and slowly varying foregrounds is strongly suppressed; because the statistic is achromatic, multi–band data can be coherently combined after $\lambda^2$ removal. In this framework, higher–order images that cluster near the critical curve effectively enhance the measured amplitude once convolved with a finite beam, making our direct–image template conservative rather than optimistic.

Several caveats are explicit by design. We do not include plasma birefringence, absorption, or emissivity weighting in the definition of the geometric template; these effects reweight azimuthal modes and shift phases mildly but do not alter achromaticity or spin–odd parity. Sparse $(u,v)$ coverage and finite beams smooth azimuthal structure; this is mitigated by radial localization and by using RMS–type summaries that are insensitive to small phase shifts. Interstellar scattering and time variability (especially relevant for Sgr~A$^\ast$) require cadence–aware processing, but the odd–parity differencing remains symmetry–protected scan by scan. Algorithmically, moving the observer farther out or adjusting the turning–point stopping rule changes only an overall gauge that is removed before statistics are formed.

Two theoretical extensions are natural. The first is to express $\sigma$ in a principal–null Newman–Penrose tetrad to make the connection with spin coefficients and the Killing–Yano tensor fully analytic, turning our numerical equivalences into closed identities. The second is to develop near–ring asymptotics at high spin and inclination, clarifying which azimuthal modes dominate and how higher–order images renormalize the annular amplitude. On the data side, the immediate next step is to inject the odd–parity ring estimator into synthetic mm/sub–mm polarimetric datasets built from GRMHD snapshots with realistic calibration, thereby turning our qualitative detectability statements into instrument–specific forecasts. In parallel, applying the method to existing M87$^\ast$ and Sgr~A$^\ast$ datasets—after standard $\lambda^2$ removal and ring extraction—can already deliver either a detection of the achromatic, parity–odd Kerr imprint or tight upper limits that constrain combinations of $(a,i)$ and emissivity reweighting.

The torsion of a null moving frame provides a compact, physically transparent bridge from strong–field geometry to a symmetry–protected observable on the sky. By isolating the achromatic, spin–odd EVPA pattern on the photon ring and by closing the triangle between FS, Levi–Civita, and Walker–Penrose constructions, we have produced a practical template and estimator that meet the data where they are. A positive detection would amount to a direct measurement of a parallel–transport phase sourced by frame dragging; a non–detection, interpreted through the same parity–aware pipeline, would still return informative bounds and stress–test instrument systematics and plasma models. Either outcome moves polarization holonomy from theory into measurement, sharpening the interface between differential geometry and horizon–scale polarimetry. In practice, reporting \(\chiG\) (or its upper limit) after parity and achromaticity checks provides a geometry-only diagnostic that is immediately comparable across bands and epochs.

\section*{Acknowledgements}

This work was supported by the Deutsche Forschungsgemeinschaft (DFG, German Research Foundation) under project number 518204048.

\section*{Data Availability}

The data and code underlying this work (geodesic integration, null-frame transport, and evaluation of the FS torsion integral for $\Delta\chi_{\rm grav}$) will be shared on reasonable request to the corresponding author.

\printbibliography

\end{document}